\documentclass[10pt]{article} 
\usepackage[utf8]{inputenc}         
\usepackage[margin=1in]{geometry}   
\usepackage{amsmath}                
\usepackage{amssymb}                
\usepackage{graphicx}               
\usepackage[T1,T2A]{fontenc}        
\usepackage[english]{babel} 

\title{\tt\bf Comparative analysis of the selected planet transits observed at the Kyiv comet station and the TESS and Kepler database.}

\author{M. Lobodenko$^1$, Ya. Pavlenko$^2$, I. Kulyk$^2$, 
A. Nahurna$^1$, M. Solomakha$^1$, O. Baransky$^3$ \\
    $^1$ Taras Shevchenko National University of Kyiv, Kyiv, Ukraine \\
    $^2$ Main Astronomical Observatory of the NAS of Ukraine, Kyiv, Ukraine\\
    $^3$ Kyiv Comet station Lisnyky, Kyiv, Ukraine}

\date{lobodenkomo.at.ukr.net}

\begin{document}

\maketitle

\begin{center}
    ABSTRACT
    \end{center}
  We present a comparative analysis of observations of the selected exoplanet transits gathered at the Kyiv comet station and those from the TESS (Transiting Exoplanet Survey Satellite) and Kepler space telescope archive. The light curves obtained by the TESS and Kepler orbital telescopes were processed using a program based on the Python package Lightkurve v2.3 which is freely available in the MUST archive (Barbara A. Mikulski Archive for Space Telescopes). The ground based observations were carried out with the 70-cm telescope AZT-8 (Lisnyky). Photometric processing of the ground based observation was performed by using the Muniwin program. The light curves and parameters of the observed transits as well as the exoplanet orbital parameters obtained from the ground based observations were published in the ETD (Exoplanet Transit Database). The determined transit parameters were compared with the results stored in the MUST archive. Here we presents a comparison of the transit periods, depths, transit durations and some planet orbital parameters obtained from two independent sets of  the observations, terrestrial and orbital, performed  at different epochs.

  \section{Introduction}
   There are a large number of methods for finding exoplanets. The transit method is one of the most effective. The planet covers part of the star when it passes over the disk of a star and the visible brightness falls (Winn, 2010). The time sequence of the observed during a transit events allow us to see these events. The magnitude of the fall in brightness depends on the relative size of the star and the planet. Therefore, the light curve provides information about the radius of the planet and some orbital parameters. But this method has several disadvantages. First, the plane of the planet's orbit should be located in a such way that we can observe the passage of the planet over the star's disk. Secondly, the planet must be large enough to  to create a detectable drop in the star brightness. For this reason most of the planets are found by the transit method and all the planets that we present in this investigation are hot Jupiters. \\
   Hot Jupiters are planets with a mass of the order of the mass of Jupiter, which rotate close to their host  stars and are always turned to them only by the one side. Therefore the period of the planet`s rotation is small, which allows us to observe transits regularly and makes hot Jupiters the most convenient targets for the observations. \\
   The TESS and Kepler orbital telescopes conducted searches for exoplanets using the transit method. The quality of observations obtained from the orbital telescopes is definitely higher than the quality of observations obtained from ground-based telescopes. We are not hindered by various atmospheric phenomena we do not depend on the weather, time of day, phase of the Moon, etc. However, the ground based observations allow us to gather the data over large time span, which provide information about possible changes in transit parameters. The best result can be achieved by combining these two types of observations. To do this, it is necessary to determine how the results obtained from space and ground-based observations correspond to each other, and whether ground observations can be considered sufficiently accurate. The main goal of our  research is to determine to what extent the parameters of exoplanets obtained from observations at the Lisnyky comet station coincide with the parameters obtained from observations from the TESS and Kepler orbital telescopes.\\[2mm]

    \section{The Kepler and TESS data bases}

    \subsection{MAST}
MAST (The Mikulski Archive for Space Telescopes) is an archive of data from numerous space telescopes and contains data obtained in the optical, ultraviolet and near-infrared ranges. The MAST archive also provides the \textbf {parameters  of the planet transits (periods, transit depths, the phases of transit events, and some planet orbital parameters) calculated by the TESS and Kepler pipelines} for all events surpassing some threshold. We used these published data in order to compare with the results of our ground based observations as well as with those transit parameters obtained from the Kepler and TESS light curves  using the program codes developed on the base of the Python package Lightkurve v2.3.

\subsection{Kepler and TESS space missions}

The Kepler Orbital Telescope was launched by NASA on March 7, 2009 to search for Earth-sized planets. The spacecraft repeated the path of the Earth, 
revolving around the Sun. This arrangement allowed telescope to constantly monitor one part of the sky. 
The field of view covered 115 square degrees near the plane of the Milky Way. The telescope of 1,039 kilograms mass contained a Schmidt camera with a 0.95-meter front corrector plate feeding a 1.4-meter primary mirror. The light reflected by the mirror  was collected in the main focus, where there was a mosaic of 21 pairs of specially created astronomical CCD matrices, capable of recording almost every incident photon. The dimensions of the entire mosaic are approximately 30 x 30 cm and it consists of 95 megapixels. In May 2013 the telescope's second flywheel engine failed. A year after the failure of the engine, the telescope began to transmit data to Earth again. The new mission has been named K2. The telescope began to observe a section of the sky along the ecliptic. Over 9 years of operation, the telescope discovered more than 2,680 exoplanets, 550 of which may be rocky, and 21 potentially habitable (Hall \& Barentsen 2020a).

The TESS (Transiting Exoplanet Survey Satellite) space telescope was launched on April 19, 2018. 
The mission was planned for two years, during these years the telescope should examine the entire area of the sky. 
The celestial sphere was divided into 26 observation sectors, each sector being 24° × 96° to detect transits of previously unknown exoplanets near the closest and brightest stars. TESS would focus on G, K, and M-type stars with apparent magnitudes brighter than magnitude 12, and 1000 nearest red dwarfs. The rotational period of the telescope is 13.7 days. Each sector is observed for 27.4 days. The sole instrument on TESS is formed of four wide-angle CCD cameras. Each camera has a 16.8-megapixel detector with a low energy consumption and low noise, which was developed in the Laboratory of Lincoln. Each camera has a 24° × 24° field of view, a 100 mm effective pupil diameter, a lens assembly with seven optical elements, and a band-pass range from 600 to 1000 nm (Hall \& Barentsen 2020a).
\\

\subsection{ The data processing}
The program we used to process data from the TESS and Kepler orbital telescopes was developed based on the Python package Lightkurve v2.3, which is available in the MUST archive.\\

First, the program finds data for the object we are interested in the catalogs (Hall \& Barentsen 2020a) and download the data. If the observed data cover several sectors, the light curves are stitched and normalized. The next step is searching for and subtracting the long-period garmonical oscillations, which can be star's oscilations and some kind of artifacts (Hall \& Barentsen 2020d, 2020e). The Lomb-Scargle periodogram is used for this(Hall \& Barentsen 2020c). To detect transit events we build a periodogram, using the Box Least Squares (BLS) method (Saunders 2020). This method is much more sensitive to find periodic transit events (Terebizh 1992). Fig. 1 presents the periodograms of star TIC236887394 from the TESS data base constructed with Lomb-Scargle and BLS methods.\\

\begin{figure}
\centering
    \includegraphics[width=12cm]{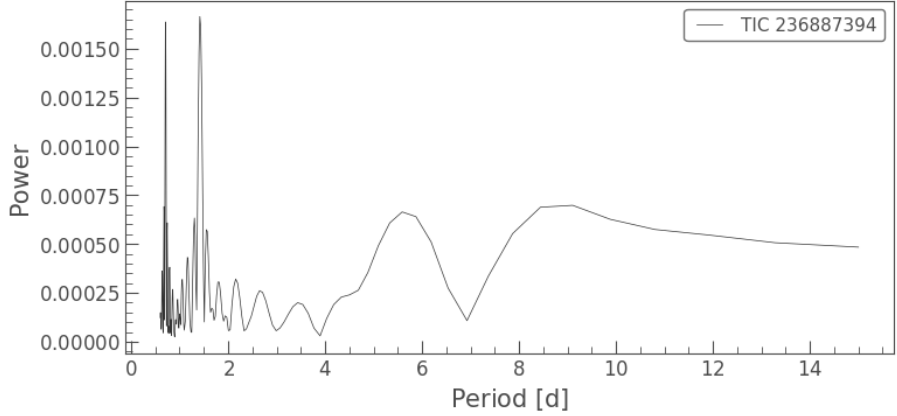}
\caption{Lomb-Scargle periodogram of star TIC 236887394 from the TESS data base}
\end{figure}
\begin{figure}
\centering
    \includegraphics[width=12cm]{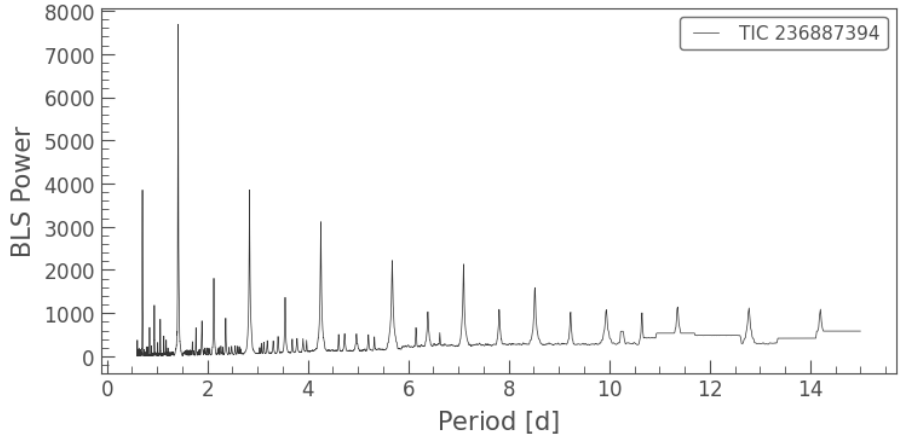}
\caption{BLS periodogram of star TIC 236887394 from the TESS data base} 
     \end{figure} 

The program outputs are period, duration, and first epoch of the detected transit based on the periodogram built with the BLS method for the period at maximum power. Using these parameters we construct the folded light curve in the phase space, which helps to notice a shifts (if any) in the moments of the beginning and end of the transit. These displacements may indicate the presence of one or more planets in this system. Fig.3 demonstrates the folded phase curve for star TIC 236887394 with the considerable decrease in the star brightness caused by the transit event. 

\begin{figure}
\centering
    \includegraphics[width=12cm]{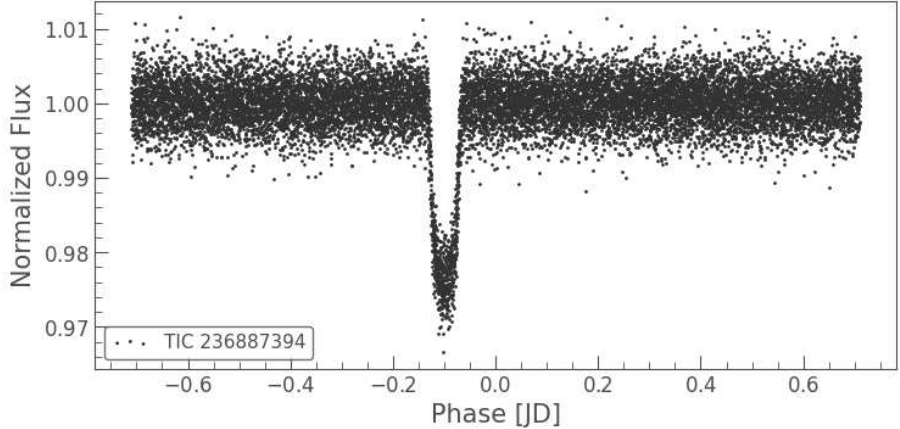}
\caption{{\bf Folded} phase curve of star TIC 236887394}
\end{figure}

In order to find other possible transits we cut out parts of the light curve where the first transit occurs and repeat the procedure again building periodogram using the BLS method.

\section{Observations obtained at the Lisnyky comet station}
\subsection{ Observations}
Our observations were also carried out from March 24, 2021, to February 14, 
2022, at the Lisnyky comet station \textbf {with} the 70-centimeter AZT-8 reflector 
telescope, using the R filter. \\
The telescope is equipped with a FLI PL4710 back illuminated CCD and UBV 
RI Bessel filters. For faint objects we use a mode with 2 x 2 binning, 
which gives a scale of 1.96 arcsec/pixel. The FoV of the instrument is 16 × 16 arcmin. The limiting magnitude of a 300 s exposure image is 20 mag under good sky conditions. It is possible to reach 21.5-22 mag with 1800 s exposures. For our observations, the exposure time varied from 10 to 30 seconds, depending on the brightness of the observed object.\\
We didn't observe one particular object all the night. We chose the star and the time span for the expected transit event and conducted observations during the transit only. We began shooting half an hour before the intended start time of the event and completed half an hour after the intended completion to identify possible displacements in the moments of the beginning and end of the transit. Table 1 gives the list of he observed objects and some parameters of their host stars.

\subsection{List of observed objects and their characteristics}
\begin{tabular}{p{1.2cm}p{1.2cm}p{0.9cm}p{1.1cm}p{1.9cm}}
\hline  
      & Spectral type & Fluxes & Distance [pc] & Constellation\\ 
     \hline
     TrES-3 & G & 12.4 & 231.34 & Hercules\\  
     
     Kepler-17 & G5V & 14.14 & 720.75 & Cygnus\\ 
     
     Qatar-1 & K & 12.65 & 185.62 & Draco\\ 
     
     WASP-43 & K7V & 12.08 & 86.75 & Sextans\\ 
     
     WASP-3 & F7V & 10.52 & 231.16 & Lyra\\ 
      
     TrES-5 & K & 13.77 & 360.31 & Cygnus\\ 
     
     Qatar-2 & K5V & 13.45 & 181.37 & Virgo\\ 
     
     WASP-10 & K5V & 12.03 & 141 & Pegasus\\ 
     
     \hline
\end{tabular}

\subsection{Processing observations with Muniwin}

The processing of the observations obtained in the Lisnyky comet station was carried out using the C-Munipack package https://c-munipack.sourceforge.net/. This program uses differential photometry: we compare the brightness of two or more stars. One star is one which brightness must change due to the transit of the planet through its disk, and the other (others) are reference stars which brightness should be unchanged. First, we calibrate the images using dark, bias, flatfield files. The next step is to set the parameters by which the program determines which object in the image is the star. The next step is to look for the corresponding stars on each image. After that, we select a star through the disk of which transit should take place and the stars which brightness should be unchanged. Reference stars cannot be variable stars. We use the Simbad database to make sure that the selected star has constant brightness. When the necessary stars are selected, we proceed to the light curve construction. 

\begin{figure}
\centering
\includegraphics[width=12cm]{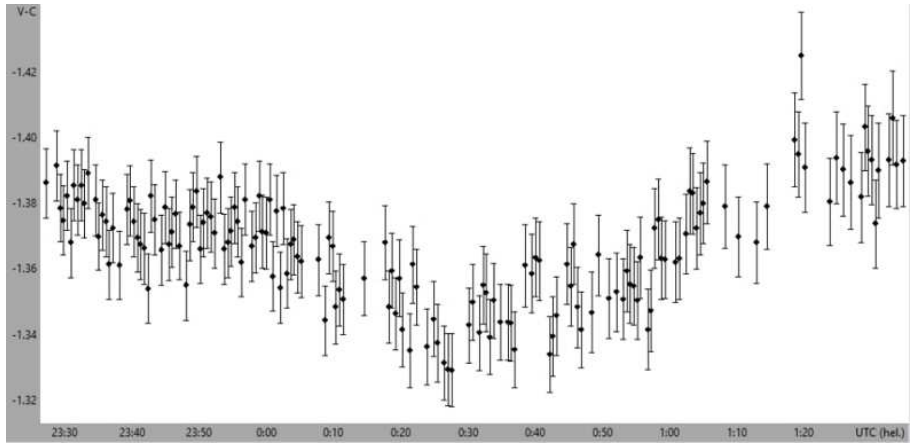}
\caption{The light curve for the star TIC 36734222 obtained using the C-Munipack package.}
    \end{figure}

The resulting light curve was uploaded to the Czech Exoplanet Transit Database (ETD) for further processing.

\subsection{Exoplanet Transit Database}

 The ETD has an algorithm for processing light curves, by which it determines the main transit parameters from these light curves: the moments of the beginning, end and middle of transit, the depth, duration and radius of the planet, and the inclination of the orbit. The period of the planet is determined from all observations published in the database.\\

All our observations were uploaded to this database and received data quality 2 and 3. The data quality in the ETD is estimated on a scale from 1 to 5, where 1 is the data with the highest quality.

\section{Results}

The parameters of the exoplanet transits obtained from the ground-based observations and determined from the data base of the orbital telescopes are listed in Table 1.\\
We could not extract some parameters from our own observations, therefore they were taken from ETD based on all observations published in this data base. Such parameters are indicated by symbol "*". The table also comprise  the periods and transit durations calculated in this work from the Kepler and TESS light curves.\\
\\
\section{Conclusion}
The comparison of the transit parameters determined from two different data sets, i.e., ground-based and space-based, indicates that: \\
1) The transit periods and durations obtained from the ground-based observations with the small telescope agree well with those obtained from Kepler and TESS orbital telescopes, highlighting the high accuracy of the ground-based observations presented in this work. \\
2) It is shown that the average agreement of transit parameters is at the level of 0.00001 day (transit period) and 0.1 hour (transit duration) \\
3) Independent processing of the light curves from the Kepler and TESS databases indicates that the preliminary processing of the light curves (the light curve detrending, removing of the long-periodic harmonic stellar oscillations) can significantly affect the accuracy of the extracted transit parameters.\\
\\

{\it Acknowledgements.} YP anf IK participation was performed in the frames of 
the government funding program for institutions of the National Academy of Sciences
of Ukraine (NASU) and supported by the National Research Foundation of
Ukraine (Project N 2020.02/0228). Part of the data presented in this
paper were obtained from the Mikulski Archive for Space Telescopes (MAST),
which is hosted at STScI. STScI is operated by the Association of Universities
for Research in Astronomy, Inc., under NASA contract NAS5-26555. Support for
MAST for non-HST data is provided by the NASA Office of Space Science via
grant NNX09AF08G and by other grants and contracts. This paper includes data
collected by the TESS mission. Funding for the TESS mission is provided by
the NASA Science Mission directorate.

\section{References}
 Bunnell J.: 2018, Using Kepler Data to Plot a Light Curve, [online] 
Available at: (https://github.com/spacete\-lescope notebooks/blob/master/notebooks/MAST/Kepler/Kepler\_Lightcurve /kepler\_lightcurve.ipynb)\\
 \\
 Bunnell J.: 2018, Plotting Images from Kepler Target Pixel Files, [online] Available at: 
(https://github.com/\\
spacetelescope notebooks/blob /master/notebooks/MAST/Kepler/Kepler\_TPF/ kepler\_tpf.ipynb)\\
\\
Hall O., Barentsen G.: 2020a, Using Kepler Light Curve Products with Lightkurve, 
[online] Available at: (https://github.com/spacetelescope/notebooks/blob
/master/notebooks/MAST/Kepler/kepler\_using\_\\light\_curve\_ products\_with\_lightkurve/kepler\_using \_light\_curve\_products\_with\_lightkurve.ipynb)\\
\\
Hall O., Barentsen G.: 2020b, Using Kepler Target Pixel File Products with 
Lightkurve, [online] Available at: 
(https://github.com/spacetelescope/notebooks/blob/master/notebooks/MAST/Kepler/
kepler\_using\_ target\_pixel\_file\_products\_with\_lightkurve/kepler\_using\_target\_pixel\_file\_products\_with\_lightkurve\\.ipynb)\\
 \\
Hall O., Barentsen G.: 2020c, Measuring and Removing a Rotation Period Signal from a Kepler Light Curve,[online] Available at: (https://github.com/spacetelescope/notebooks/blob /master/notebooks/MAST\\/Kepler/kepler\_measuring \_a\_rotation\_period/kepler\_measuring\_a\_rotation \_period.ipynb)\\
\\
Hall O., Barentsen G.: 2020d, Creating Periodograms and Identifying Significant Peaks, [online] Available at: (https://github.com/spacetelescope/notebooks/blob /master/notebooks/MAST/Kepler/kepler\_creating\_\\periodograms/kepler\_creating\_periodograms.ipynb)\\
\\
Hall O., Barentsen G.: 2020e, How to Understand and Manipulate the Periodogram of an Oscillating Star, [online] Available at: (https://github.com/spacetelescope/notebooks/blob /master/notebooks/MAST/Kepler/\\kepler\_how\_to\_ understand\_and\_manipulate\_the\_periodogram\_of \_an\_oscillating\_star/kepler\_how\\\_to\_understand\_ and\_manipulate\_the\_periodogram\_of\_an\_oscillating \_star.ipynb)\\
\\
Pavlenko Ya., Shubina O., Kulyk I., Kuznyetsova Y., Zakhozhay O., Korsun P., Borysenko S., Krushevska V., Andreev M.: 2021, Exocometary Activity Around Stars at Different Evolutionary Stages: Current Issues (Main Astronomical Observatory of the NAS of Ukraine, Kyiv, Ukraine)\\
\\
 Pavlenko, Ya., I. Kulyk, O. Shubina, M. Vasylenko, D. Dobrycheva, P. Korsun: 2022 (arXiv:2202.13373)
\\ 
  Saunders N.: 2020, Identifying Transiting Planet Signals in a Kepler Light Curve, [online] Available at: (https://github.com/spacetelescope/notebooks/blob
 /master/notebooks/MAST/Kepler/kepler\_identifying \_transiting\_planet\_signals/kepler\_identifying\_ transiting\_planet\_signals.ipynb)\\
\\
 Terebizh V.Y.: 1992, Analysis of time series in astrophysics (Moscow "Nauka" Gl.ed.fiz.-mat.lit.)
\\\\ 
Winn, J.~N., 2010, Exoplanet Transits and Occultations, in "Exoplanets", ed. Seager, S., pp. 55-77. 

\onecolumn
    {\bf Table 1.} Parameters of the selected transits obtained from the ground based and orbital telescopes.\\
    \\
{\bf Kepler 17b}\\
\begin{tabular}{p{22mm}p{1.5cm}p{1.2cm}p{2.2cm}}
\hline  
      & Observation & MAST & obtained with Lightkurve 2.0 \\ 
     \hline
     Period [day] & 1.4857108* & 1.4857108 & 1.4856924 \\
     Transit duration [hour] & 2.41 & 2.29 & 2.40 \\
     Depth & 0.0201 [mag] & 1.7765[\%] &  \\
     Planet radius $[R_j]$ & 1.34 & 1.31 &  \\
     Inclination [$^\circ$] & 90.00 & 87.2 &  \\
     R$_{\rm orbit}$ [AU] & 0.026* & 0.026 &  \\
     \hline
\end{tabular}\\
\\
\\
\\
{\bf TrES-3b}\\
\begin{tabular}{p{3cm}p{1.7cm}p{1.4cm}p{2.2cm}}
\hline  
      & Observation & MAST & obtained with Lightkurve 2.0 \\ 
     \hline
     Period [day] & 1.3061861* & 1.3061858 & 1.3061269 \\
     
     Transit duration [hour] & 1.31 & 1.37 & 1.2 \\
     
     Depth & 0.0279 [mag] & 2.7361[\%] &  \\
     
     Planet radius $[R_j]$ & 1.262 & 1.336 &   \\
     
     Inclination [$^\circ$] & 81.76 & 81.93 &  \\
     
     R$_{\rm orbit}$ [AU] & 0.023* & 0.023 &  \\
     \hline
\end{tabular}\\
\\
{\bf Qatar-1b}\\
\begin{tabular}{p{3cm}p{1.7cm}p{1.4cm}p{2.2cm}}
\hline  
      & Observation & MAST & obtained with Lightkurve 2.0 \\ 
     \hline
     Period [day] & 1.4200246* & 1.4200242 & 1.4202261 \\
     
     Transit duration [hour] & 1.59 & 1.66 & 1.37 \\
     
     Depth & 0.0203 [mag] & 2.1557[\%] &  \\
     
     Planet radius $[R_j]$ & 1.093 & 1.143 &   \\
     
     Inclination [$^\circ$] & 83.43 & 84.48 &  \\
     
     R$_{\rm orbit}$ [AU] & 0.023* & 0.023 &  \\
     \hline
\end{tabular}\\
\\
{\bf WASP-43b}\\
\begin{tabular}{p{3cm}p{1.7cm}p{1.4cm}p{2.2cm}}
\hline  
      & Observation & MAST & obtained with Lightkurve 2.0 \\ 
     \hline
     Period [day] & 0.81347414* & 0.813475 & 0.8134242 \\
     
     Transit duration [hour] & 1.05 & 1.16 & 1.2 \\
     
     Depth & 0.0280 [mag] & 2.5371[\%] &   \\
     
     Planet radius $[R_j]$ & 0.93 & 0.93 &   \\
     
     Inclination [$^\circ$] & 81.18 & 82.15 &  \\
     
     R$_{\rm orbit}$ [AU] & 0.014* & 0.014 &  \\
     
     \hline
\end{tabular}\\
\\\\
\\\\
{\bf WASP-3b}\\
\begin{tabular}{p{3cm}p{1.7cm}p{1.4cm}p{2.2cm}}
\hline  
      & Observation & MAST & obtained with Lightkurve 2.0 \\ 
     \hline
     Period [day] & 1.846835* & 1.846830 & 1.846529 \\
     
     Transit duration [hour] & 2.22 & 2.79 & 2.40 \\
     
     Depth &  0.0124[mag] & 1.0269[\%] &   \\
     
     Planet radius $[R_j]$ & 1.36 & 1.42 &   \\
     
     Inclination [$^\circ$] & 81.71 & 84.15 &  \\
     
     R$_{\rm orbit}$ [AU] & 0.0317* & 0.032 &  \\
     
     \hline
\end{tabular}\\
\\
{\bf TrES-5b} \\
\begin{tabular}{p{3cm}p{1.7cm}p{1.4cm}p{2.2cm}}
\hline  
      & Observation & MAST & obtained with Lightkurve 2.0 \\ 
     \hline
     Period [day] & 1.48224718* & 1.48224690 & 1.48225292 \\
     
     Transit duration [hour] & 1.852 & 1.846 & 1.200 \\
     
     Depth & 0.0224 [mag] & 2.192 [\%] &  \\
     
     Planet radius $[R_j]$ & 1.208 & 1.256 &   \\
     
     Inclination [$^\circ$] & 84.54 & 84.53 &  \\
     
     R$_{\rm orbit}$ [AU] & 0.025* & 0.025  &  \\
     
     \hline
\end{tabular}\\
\newpage
\end{document}